\documentclass[a4paper,11pt]{article}
\usepackage{pos}
\usepackage{subcaption}
\title{Hadronic Parity Violation from 4-quark Interactions}

\author*[a]{Aniket Sen}
\author[a]{Marcus Petschlies}
\author[a]{Nikolas Schlage}
\author[a]{Carsten Urbach}

\affiliation[a]{Helmholtz-Institut für Strahlen- und Kernphysik, Rheinische Friedrich-Wilhelms-Universität Bonn,\\
  Nussallee 14-16, 53115 Bonn, Germany}

\emailAdd{sen@hiskp.uni-bonn.de}

\newcommand{\fermi}{\,\mathrm{fm}}
\newcommand{\gev}{\,\mathrm{GeV}}
\newcommand{\mev}{\,\mathrm{MeV}}
\newcommand{\xvec}{\vec{x}}

\abstract{We present an exploratory investigation of the parity odd
  $\Delta I = 1$ pion-nucleon coupling $h_\pi^1$ from lattice
  QCD. Based on  the PCAC relation, we study the parity-conserving
  effective Hamiltonian and extract the coupling by determining the
  nucleon mass splitting arising from the effective 4-quark
  interactions using the Feynman-Hellmann theorem.  
  We present preliminary results of the mass shift for a $32^3 \times
  64$ ensemble of $N_f = 2 + 1 + 1$ twisted mass fermions at pion 
  mass $260 \mev$ and lattice spacing $a = 0.097 \fermi$.
}

\FullConference{%
 The 38th International Symposium on Lattice Field Theory, LATTICE2021
  26th-30th July, 2021
  Zoom/Gather@Massachusetts Institute of Technology
}

\begin{document}
\maketitle

\section{Introduction}

The study of parity (P) violation in the weak sector of the Standard Model started in the 1960s, immediately following the seminal paper by Lee and Yang~\cite{LY} and the first experimental evidence by Madam Wu and collaborators~\cite{Wu}. Despite countless experiments over the last six decades, the study of hadronic parity violation (HPV) is still of importance. Our focus here is on flavor-conserving 
hadronic interactions. For the latter there is an overwhelming background from strong interaction,
governed by quantum chromodynamics (QCD), and electromagnetic interaction. 
In particular, since the strong coupling is significantly stronger than the weak coupling,
extracting the smaller signal contribution from P-violating hadronic weak interaction is rendered a difficult task. 

The most used theory model of HPV is based on the exchange of single light mesons, as developed by Desplanques, Donoghue and Holstein (DDH)~\cite{DDH}. This model contains seven independent coupling constants. 
A long-standing problem in HPV is the determination of these P-violating coupling constants. Of particular interest to us is the $\Delta I = 1$ pion-nucleon coupling $h_\pi^1$. This particular transition is suppressed for the weak charged current, 
and therefore serves as a probe for the in this case dominant neutral current.
Since the underlying weak operators and the Wilson coefficients are known,
the only remaining work is the calculation of the hadronic matrix elements, either through non-perturbative methods or through fits to experimental data.

On the side of experiments, a recent measurement by the NPDGamma collaboration produced the first experimental determination of $h_\pi^1$~\cite{NPDGamma}. On the theoretical side, however, there has been only one study using lattice QCD~\cite{Wasem}. Despite the fact that this theoretical determination overlaps very nicely with the experimental measurement, one needs to be careful when considering this result. The study was done on a single lattice with pion mass of 390 MeV,
and contributions from the so-called "quark-loop" diagrams were not considered.
The author also used 3-quark interpolators for the $N \pi$ final state in order to simplify Wick contractions. 
These issues were pointed out by Feng, Guo and Seng (FGS) in their 2018 paper~\cite{FGS}, where they also proposed a different strategy for determining $h_\pi^1$. They proposed to get rid of the $N \pi$ final state by using partially conserved axial current (PCAC) relations
and to determine the coupling constant in terms of the induced relative mass shift between proton and neutron.
Our work serves as a benchmark for this newly proposed technique.

\section{The 4-quark operators}

According to the DDH model, the parity violating weak Lagrangian, in the hadronic sector, is written as
\begin{equation}
  \begin{aligned}
    \mathcal{L}_{PV}^w = \frac{h_\pi^1}{\sqrt{2}} \, \bar{N} \, \left( \vec{\tau} \times \vec{\pi} \right)_3 \, N + \cdots\,.
  \end{aligned}
\end{equation}
Here $N = ( p \; n )^T$ is the isospin doublet of nucleons 
and $\vec{\tau}$ are the Pauli matrices. One then obtains the coupling constant $h_\pi^1$ from the hadronic matrix element
\begin{equation}
  \begin{aligned}
    h_\pi^1 = - \frac{i}{2 m_N} \lim_{p_\pi \to 0} \langle n \pi^+ \vert \mathcal{L}_{PV}^w (0) \vert p \rangle \,,
  \end{aligned}
\end{equation}
where $m_N$ is the average nucleon mass. At low energies
$\mathcal{L}_{PV}^w$ can be expressed in terms of seven independent P-odd 4-quark operators~\cite{KS}
\begin{equation}
  \begin{aligned}
    \mathcal{L}_{PV}^w = - \frac{G_F}{\sqrt{2}} \, \frac{\sin^2 \theta_w}{3} \sum_i \left( C_i^{(1)} \theta_i + S_i^{(1)} \theta_i^{(s)} \right)\,,
  \end{aligned}
\end{equation}
where
\begin{equation}
  \begin{aligned}
    \theta_1 &= \bar{q_a} \gamma_\mu q_a \bar{q}_b \gamma_\mu \gamma_5 \tau_3 q_b, \;\;\;\;\;\; \theta_2 = \bar{q}_a \gamma_\mu q_b \bar{q}_b \gamma_\mu \gamma_5 \tau_3 q_a,\\
    \theta_3 &= \bar{q}_a \gamma_\mu \gamma_5 q_a \bar{q}_b \gamma_\mu \tau_3 q_b, \;\;\;\;\;  \\
    \theta_1^{(s)} &= \bar{s}_a \gamma_\mu s_a \bar{q}_b \gamma_\mu \gamma_5 \tau_3 q_b, \;\;\;\;\; \theta_2^{(s)} = \bar{s}_a \gamma_\mu s_b \bar{q}_b \gamma_\mu \gamma_5 \tau_3 q_a\,,\\
    \theta_3^{(s)} &= \bar{s}_a \gamma_\mu \gamma_5 s_a \bar{q}_b \gamma_\mu \tau_3 q_b, \;\;\;\;\; \theta_4^{(s)} = \bar{s}_a \gamma_\mu \gamma_5 s_b \bar{q}_b \gamma_\mu \tau_3 q_a\,.
  \end{aligned}
\end{equation}
Here, $q = ( u \; d )^T$ is the light quark doublet, $G_F$ is the Fermi constant and $\theta_w$ is the weak mixing angle. $C_i$ and $S_i$ are well known Wilson coefficients. The values of these coefficients, at the scale $\Lambda_\chi \approx 1 \gev$, are~\cite{Tiburzi:2012hx}
\begin{equation}
  \begin{aligned}
    C^{(1)} (\Lambda_\chi) &= \begin{pmatrix} -0.055, & 0.810, & -0.627 \end{pmatrix}\\
    S^{(1)} (\Lambda_\chi) &= \begin{pmatrix} 5.09, & -2.55, & 4.51, & -3.36 \end{pmatrix}\\
  \end{aligned}
  \label{eq:Wilson_coeff}
\end{equation}
The presence of a soft pion in the final state of the matrix element greatly complicates the calculation for reasons pointed out in \cite{GS}. FGS proposed to get rid of the soft pion by relating this matrix element to a parity conserving counterpart, using the PCAC relation
\begin{equation}
  \begin{aligned}
    \lim_{p_\pi \to 0} \langle a \pi^i \vert \hat{O} \vert b \rangle = \frac{i}{F_\pi} \langle a \vert \left[ \hat{O}, \hat{Q}_A^i \right] \vert b \rangle\,,
  \end{aligned}
\end{equation}
where $F_\pi = 92.1 \mev$ is the pion decay constant and $\hat{Q}_A^i$ are the axial charges. This means instead of calculating $\langle n \pi^+ \vert \mathcal{L}_{PV}^w \vert p \rangle$, we can simply compute $\langle n \vert \left[ \mathcal{L}_{PV}^w, \hat{Q}_A^i \right] \vert p \rangle$. At the operator level, the PCAC relation yields
\begin{equation}
  \begin{aligned}
    \left[ \theta_q, \hat{Q}^i_A \right] = i \epsilon^{3ij} \theta'_{q (j)} , \;\;\;\;\; \left[\theta_q^{(s)}, \hat{Q}^i_A \right] = i \epsilon^{3ij} \theta_{q (j)}^{(s)'}\,,
  \end{aligned}
\end{equation}
where $\theta'_q$ and $\theta^{(s)'}_q$ are the parity even 4 quark operators defined as
\begin{align}
  \theta^{'\phantom{(s)}}_1 &= \bar{q}_a \gamma_\mu q_a \bar{q}_b \gamma_\mu \tau_3 q_b\,, \quad &\theta^{'\phantom{s}}_2 &= \bar{q}_a \gamma_\mu q_b \bar{q}_b \gamma_\mu \tau_3 q_a \,,
    \nonumber \\
     \theta^{'\phantom{(s)}}_3 &= \bar{q}_a \gamma_\mu \gamma_5 q_a \bar{q}_b \gamma_\mu \gamma_5 \tau_3 q_b\,,
    \label{eq:op4q-l} \\
    \theta_1^{(s)'} &= \bar{s}_a \gamma_\mu s_a \bar{q}_b \gamma_\mu \tau_3 q_b \,, \quad  &\theta_2^{(s)'} &= \bar{s}_a \gamma_\mu s_b \bar{q}_b \gamma_\mu \tau_3 q_a \,,
    \nonumber \\
    \theta_3^{(s)'} &= \bar{s}_a \gamma_\mu \gamma_5 s_a \bar{q}_b \gamma_\mu \gamma_5 \tau_3 q_b \,, \quad  &\theta_4^{(s)'} &= \bar{s}_a \gamma_\mu \gamma_5 s_b \bar{q}_b \gamma_\mu \gamma_5 \tau_3 q_a \,.
    \label{eq:op4q-s}
\end{align}

This maps the matrix elements of P-odd hadronic operators to matrix elements to P-even ones to leading order in chiral perturbation theory
\begin{equation}
  \begin{aligned}
    \lim_{p_\pi \to 0} \langle n \pi^+ \vert \mathcal{L}_{PV}^w (0) \vert p \rangle \approx -\frac{\sqrt{2} i}{F_\pi} \langle p \vert \mathcal{L}_{PC}^w (0) \vert p \rangle = \frac{\sqrt{2} i}{F_\pi} \langle n \vert \mathcal{L}_{PC}^w (0) \vert n \rangle\,,
  \end{aligned}
\end{equation}
where $\mathcal{L}_{PC}^w$ is the following auxiliary parity conserving Lagrangian
\begin{equation}
  \begin{aligned}
    \mathcal{L}_{PC}^w = - \frac{G_F}{\sqrt{2}} \frac{\sin^2 \theta_W}{3} \sum_i \left(C_i^{(1)} \theta'_i + S^{(1)}_i \theta^{(s)'}_i \right)\,.
  \end{aligned}
\end{equation}
Since the PCAC relation holds at the operator level, the Wilson coefficients are identical to those in $\mathcal{L}_{PV}^w$. The matrix element can be determined in terms of the neutron-proton mass splitting $(\delta m_N)_{4q} \equiv (m_n - m_p)_{4q}$ induced by the Lagrangian $\mathcal{L}_{PC}^w$
\begin{equation}
  \begin{aligned}
    (\delta m_N)_{4q} = \frac{1}{m_N} \langle p \vert \mathcal{L}_{PC}^w (0) \vert p \rangle = - \frac{1}{m_N} \langle n \vert \mathcal{L}_{PC}^w (0) \vert n \rangle\,.
  \end{aligned}
\end{equation}
Combining everything, one obtains the coupling constant $h_\pi^1$ as
\begin{equation}
  \begin{aligned}
    h_\pi^1 \approx - \frac{(\delta m_N)_{4q}}{\sqrt{2} F_\pi}\,.
  \end{aligned}
\end{equation}

\section{Mass shift from Feynman-Hellmann theorem}

According to Feynman-Hellmann theorem (FHT) \cite{FHT1,FHT2}, a perturbation in the Hamiltonian of the form $H \to H + \lambda H_\lambda$ corresponds to a variation in the spectrum as
\begin{equation}
  \begin{aligned}
    \frac{\partial E_n}{\partial \lambda} = \langle n \vert H_\lambda \vert n \rangle\,.
  \end{aligned}
\end{equation}
This straightforward relation of first order in perturbation theory can be used to extract matrix elements in lattice QCD~\cite{Bouchard}. Considering the 2-point nucleon-nucleon correlator in the presence of an external source $\lambda$
\begin{equation}
  \begin{aligned}
    C_\lambda (t) &= \langle \lambda \vert N(t) \bar{N}(0) \vert \lambda \rangle 
    = \frac{1}{Z_\lambda} \int \mathcal{D}\psi\, \mathcal{D}\bar{\psi}\; N(t) \bar{N}(0)\; e^{- S - S_\lambda}\,,
  \end{aligned}
\end{equation}
where the source $\lambda$ is coupled to the 4-quark current introduced in the previous section
\begin{equation}
  \begin{aligned}
    S_\lambda = \lambda \int d^4x\; \mathcal{L}_{PC}^w (x)\,.
  \end{aligned}
\end{equation}
The derivative of the perturbed correlator gives
\begin{equation}
  \begin{aligned}
    - \frac{\partial C_\lambda}{\partial \lambda} \bigg\vert_{\lambda = 0} &= \frac{\partial Z_\lambda}{\partial \lambda} \bigg\vert_{\lambda = 0} \frac{C(t)}{Z} + \frac{1}{Z} \int \mathcal{D} \psi \mathcal{D} \bar{\psi} e^{-S} \int d^4x\; \mathcal{L}_{PC}^w (x) N(t) \bar{N}(0) \\
    &= - C(t) \int d^4x\; \langle \Omega \vert \mathcal{L}_{PC}^w (x) \vert \Omega \rangle + \int d^4x\; \langle \Omega \vert \mathcal{T}\{N(t) \mathcal{L}_{PC}^w (x) \bar{N}(0) \} \vert \Omega \rangle\,.
  \end{aligned}
\end{equation}
The first term vanishes due to lattice symmetries. The second term is effectively $\sum_x \langle p \vert \mathcal{L}_{PC}^w (x) \vert p \rangle$. The energy shift is then obtained from the linear response of the effective mass
\begin{equation}
  \begin{aligned}
    \frac{\partial m_N}{\partial \lambda} \bigg\vert_{\lambda = 0} (t, \tau) = \frac{1}{\tau} \left[ \frac{\partial_\lambda C_\lambda (t)}{C (t)} - \frac{\partial_\lambda C_\lambda (t + \tau)}{C (t + \tau)} \right]\,.
  \end{aligned}
  \label{eq:fht}
\end{equation}
Here $C$ is the unperturbed correlator ($\lambda = 0$). We compute the right-hand side of Eq.~(\ref{eq:fht}), which is proportional to the 
mass shift $(\delta m_N)_{4q}$.

\section{Lattice Setup}

\begin{figure}[h]
    \begin{subfigure}{.33\textwidth}
        \centering
        \includegraphics[width=1\linewidth]{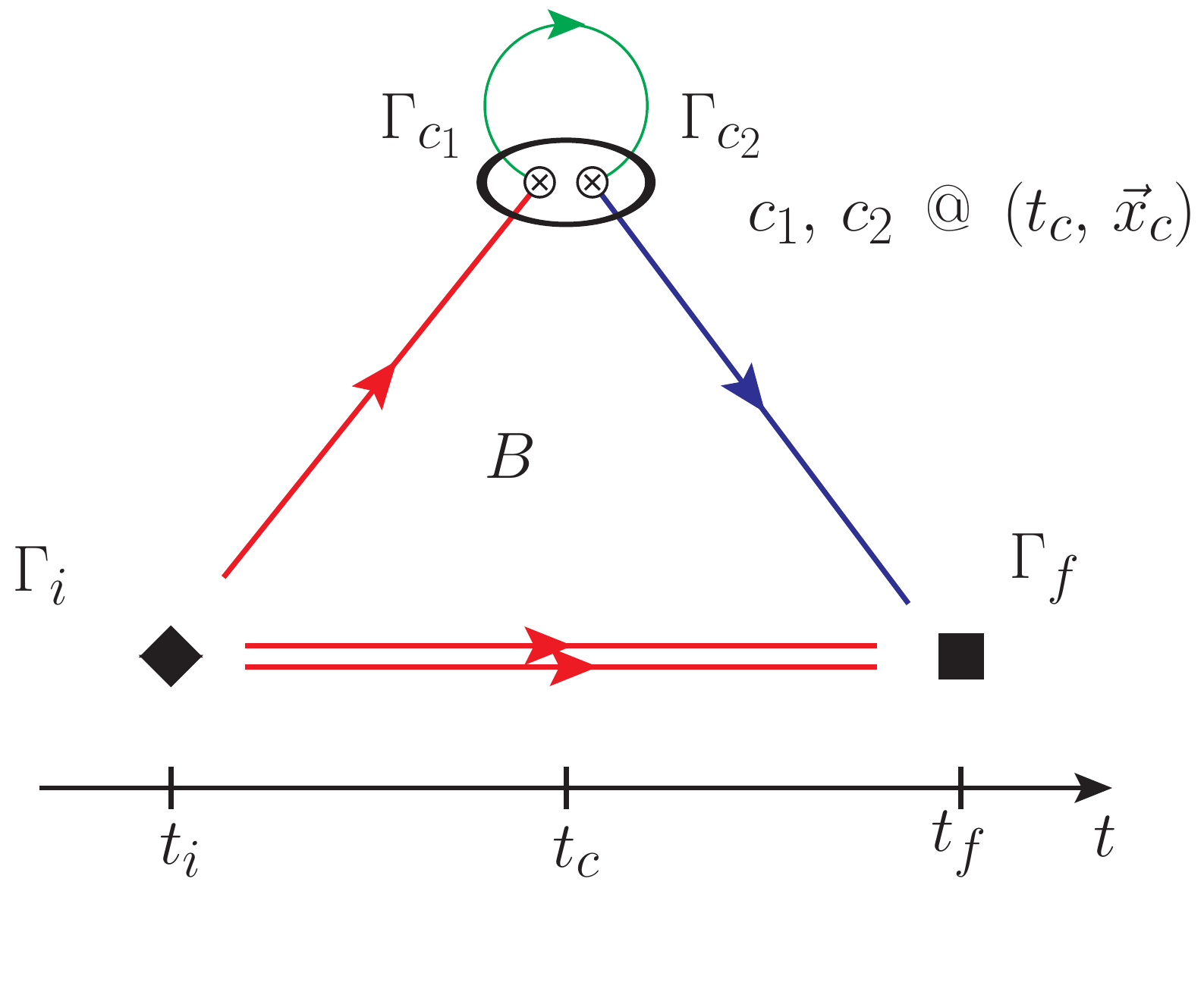}
        \caption{B-type}
    \end{subfigure}%
    \begin{subfigure}{.33\textwidth}
        \centering
        \includegraphics[width=1\linewidth]{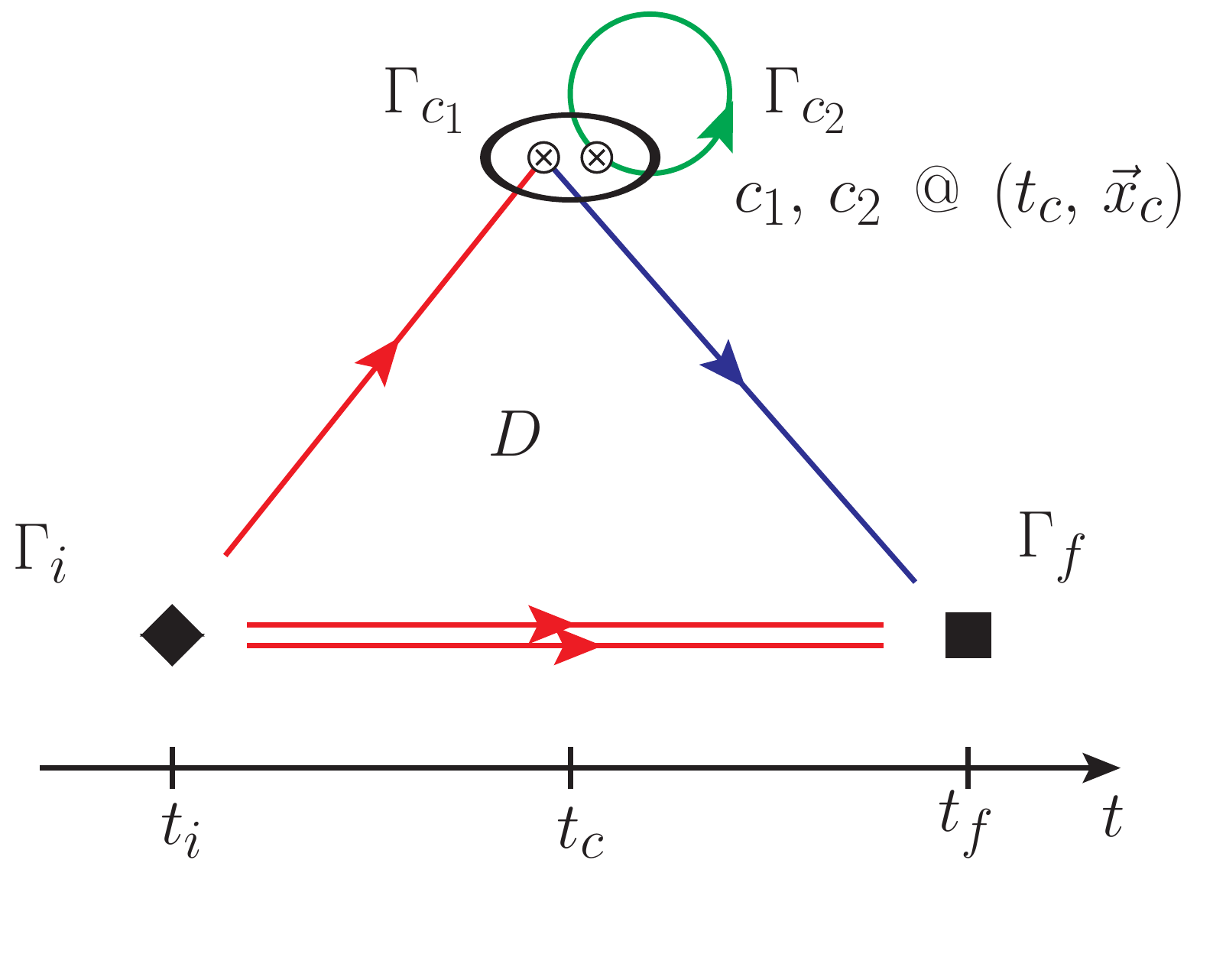}
        \caption{D-type}
    \end{subfigure}%
    \begin{subfigure}{.33\textwidth}
        \centering
        \includegraphics[width=1\linewidth]{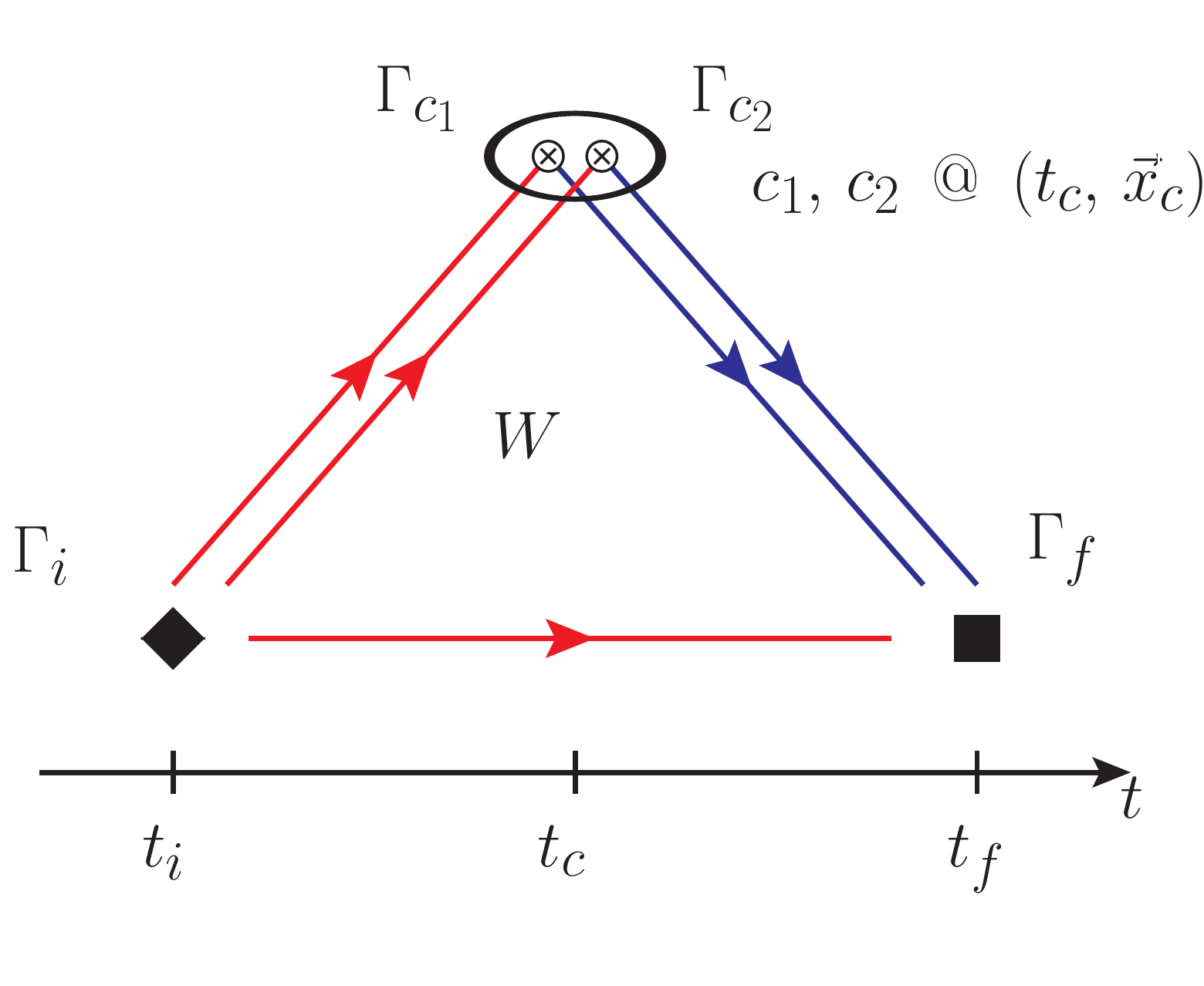}
        \caption{W-type}
    \end{subfigure}
    \caption{The 3 kinds of diagrams under consideration. (Note that $c_1$ and $c_2$ shown separately for clarity. Physically they are the same point on the lattice.) The color coding for quark propagator types is as follows: red = point-to-all, green = stochastic and blue = sequential propagator.}
    \label{fig:diagrams}
\end{figure}

The determination of $\langle p \vert \mathcal{L}_{PC}^w \vert p \rangle$ 
requires calculation of three kinds of diagrams. 
Figure~\ref{fig:diagrams} gives a visual representation of the 3 diagrams. 
There are also quark-disconnected diagrams, which vanish in the limit
of exact $SU(2)$ flavor symmetry and are neglected in the present approach.

We use sequential propagator techniques to calculate these diagrams. For B and D-type, we consider one fully time-, spin-, and color-diluted 
stochastic source $\xi^{(k)}$, where $k$ denotes the composite dilution index, 
\begin{align}
  \xi^{(t,\beta,b)}(x)_{\alpha \atop a} &= \delta_{\alpha \beta}\,\delta_{a b}\,\delta_{x_0 t}\,\xi(\xvec) \,,
  \label{eq:dilution} \\
  \xi(\xvec) &\sim \frac{1}{\sqrt{2}} \,\left( Z_2 + i Z_2 \right) \quad \mathrm{~iid} \,.
\end{align}
With the corresponding propagator $\phi_f^{(k)} = D_f^{-1}\,\xi^{(k)}$ for quark flavor $f$ 
we use the stochastic representation of the $x_c \to x_c $ quark propagator
\begin{equation}
  \begin{aligned}
    L_f(x_c)_{\alpha\beta \atop ab} = \sum\limits_{k}\,\phi_f^{(k)}(x_c)_{\alpha \atop a} \xi^{(k)*} (x_c)_{\beta \atop b}
  \end{aligned}
\end{equation}
such that $E \left[ L_f(x_c) \right] = S_f(x_c;x_c)_{\alpha\beta \atop ab}$, i.e. the stochastic expectation behaves like a quark loop.


Sequential sources for B and D-type diagrams are then built
accordingly as
\begin{align}
  \Phi_B(x_c;x_i) &:= \Gamma_{c_2}\, \bar{L}(x_c)\, \Gamma_{c_1}\, S(x_c;x_i)\\
  \Phi_D(x_c;x_i) &:= \text{Tr} \left[\Gamma_{c_2}\, \bar{L}(x_c) \right]\, \Gamma_{c_1}\, S(x_c;x_i)\,.
\end{align}
For W-type diagrams we consider a set of scalar noise sources with the properties
\begin{equation}
  \begin{aligned}
    E \left[ \eta^{(k)}(x) \right] = 0 , \hspace{2em} E \left[ \eta^{(k)}(x)\, \eta^{(l)}(y) \right] = \delta_{kl}\, \delta_{x,y}\,,
  \end{aligned}
\end{equation}
with $E$ the expectation value,
and insert such a noise between a pair of propagators to obtain a W-type propagator
\begin{equation}
  \begin{aligned}
    \tilde{W}^{(k)} (\Gamma_c) = \sum_{x_c} S(x_f;x_c)\, \Gamma_c\, \eta^{(k)}(x_c)\, S(x_c;x_i)\,.
  \end{aligned}
\end{equation}
The expectation of a product of two such propagators is needed for the W diagram.

For the simulation, we have used $N_f = 2 + 1 + 1$ clover-improved twisted mass lattice ensemble $cA211.30.32$ produced by the \emph{ETMC} collaboration \cite{config}. Table~\ref{tab:ensemble} lists the parameters of the gauge ensemble.

\begin{table}[h]
    \centering
    \begin{tabular}{|c|c|c|c|c|c|c|}
        \hline
        $L^3 \times T$ & $a [fm]$ & $a \mu_l$ & $a m_\pi$ & $m_\pi L$ & $m_\pi [MeV]$ \\
        \hline \hline
        $32^3 \times 64$ & $0.097$ & $0.00300$ & $0.12530 \, (16)$ & $4.01$ & $261.1 \, (1.1)$ \\
        \hline
    \end{tabular}
    \caption{Relevant parameters of gauge ensemble $cA211.30.32$. For
      more details we refer to Ref.~\cite{config}.}
    \label{tab:ensemble}
\end{table}

The propagators are source and sink smeared with Wuppertal smearing. APE smearing is also performed to smooth the gauge.

For the loops in B and D-type diagrams we were limited to only one
stochastic source due to computational cost. We tried overcoming this
by calculating at 8 different source coordinates per
configuration. For W-type diagrams we used 8 stochastic sources and 2
source coordinates per configuration. 

\section{Results}

\begin{figure}[h]
  \centering
  \includegraphics[width=0.7\linewidth]{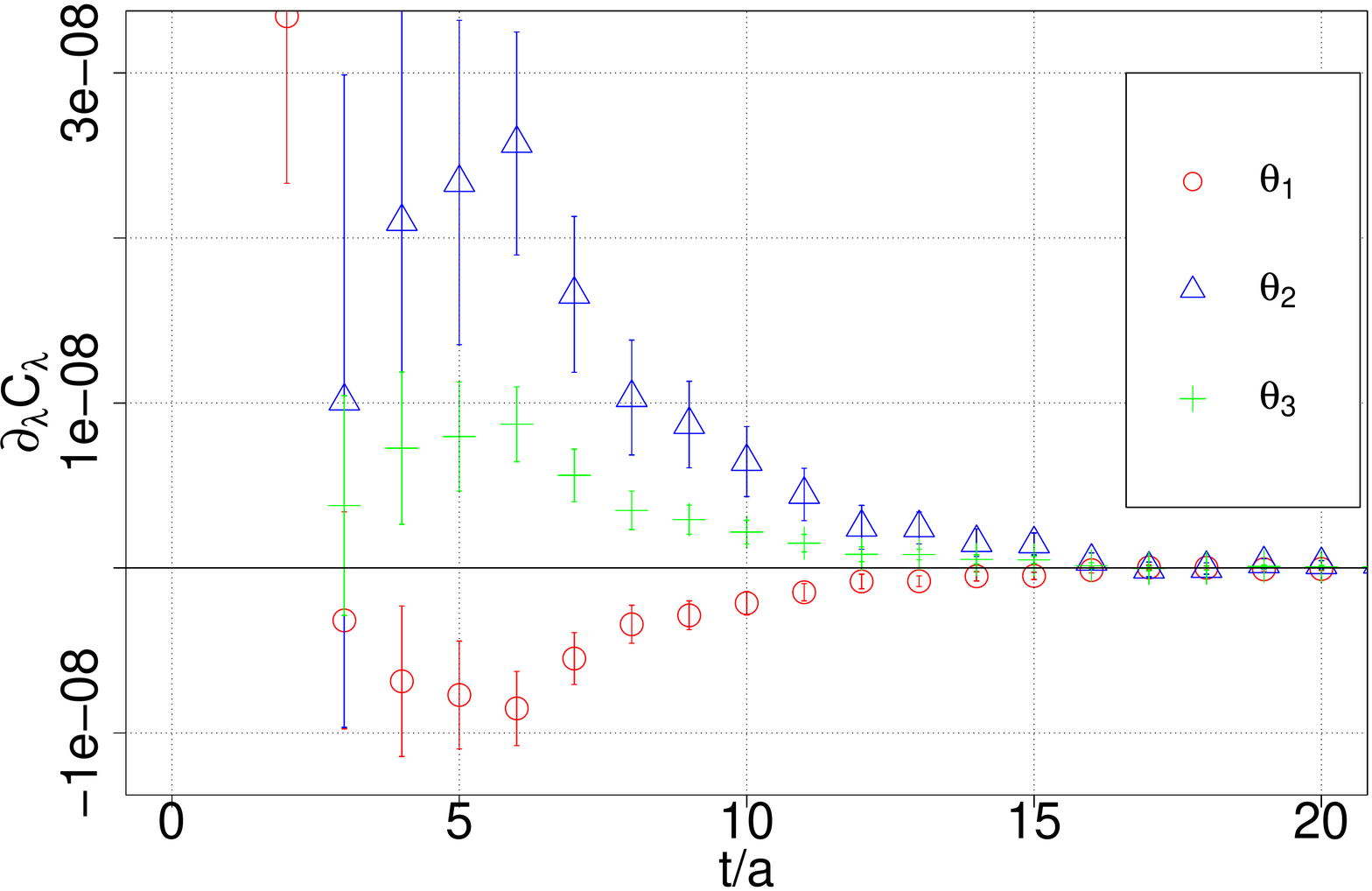}
  \caption{3-point correlators with the 4-quark operator insertions}
  \label{fig:3pt_corrs}
\end{figure}

In this work we present results for the 3 operators in the light quark
sector, Eq.~(\ref{eq:op4q-l}). Figure~\ref{fig:3pt_corrs} shows the
3-point correlators for the 3 operator insertions (note we have
dropped the $'$ in the operator names). The symmetry between
$\theta_1$ and $\theta_3$  is expected at least at tree
level. $\theta_2$ has considerably more noise. We suspect that it
could be due to mixing under renormalization with the lower
dimensional operator $\bar{\psi}\,\gamma_5\,\psi$.

\begin{figure}[h]
  \centering
  \includegraphics[width=0.7\linewidth]{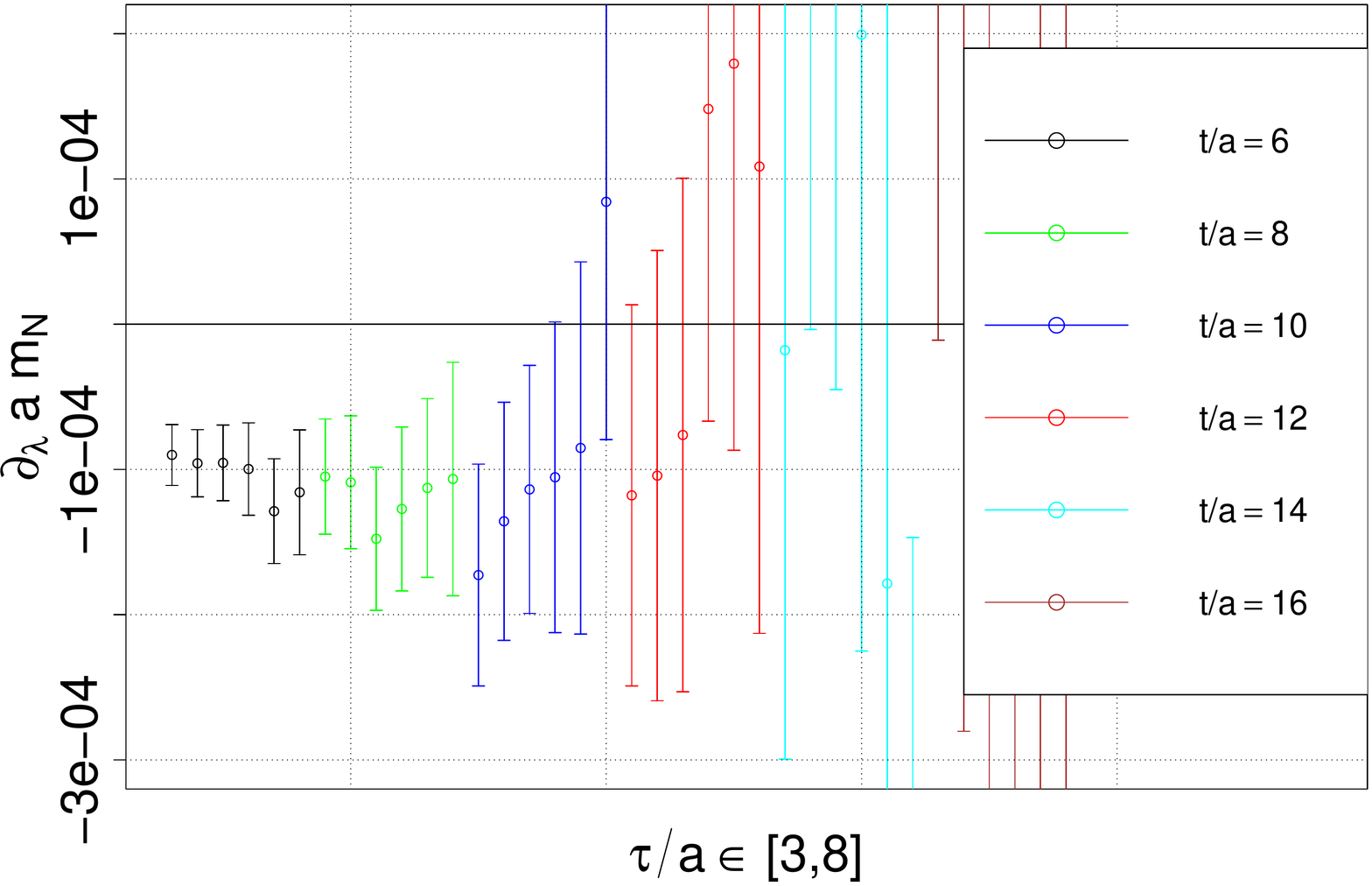}
  \caption{Large $t$ and $\tau$ dependence of $\theta_1$ (for each timeslice the data is shown for the range $\tau/a \in [3,8]$ )}
  \label{fig:theta1_t_tau}
\end{figure}

\begin{figure}[h]
  \centering
  \includegraphics[width=0.7\linewidth]{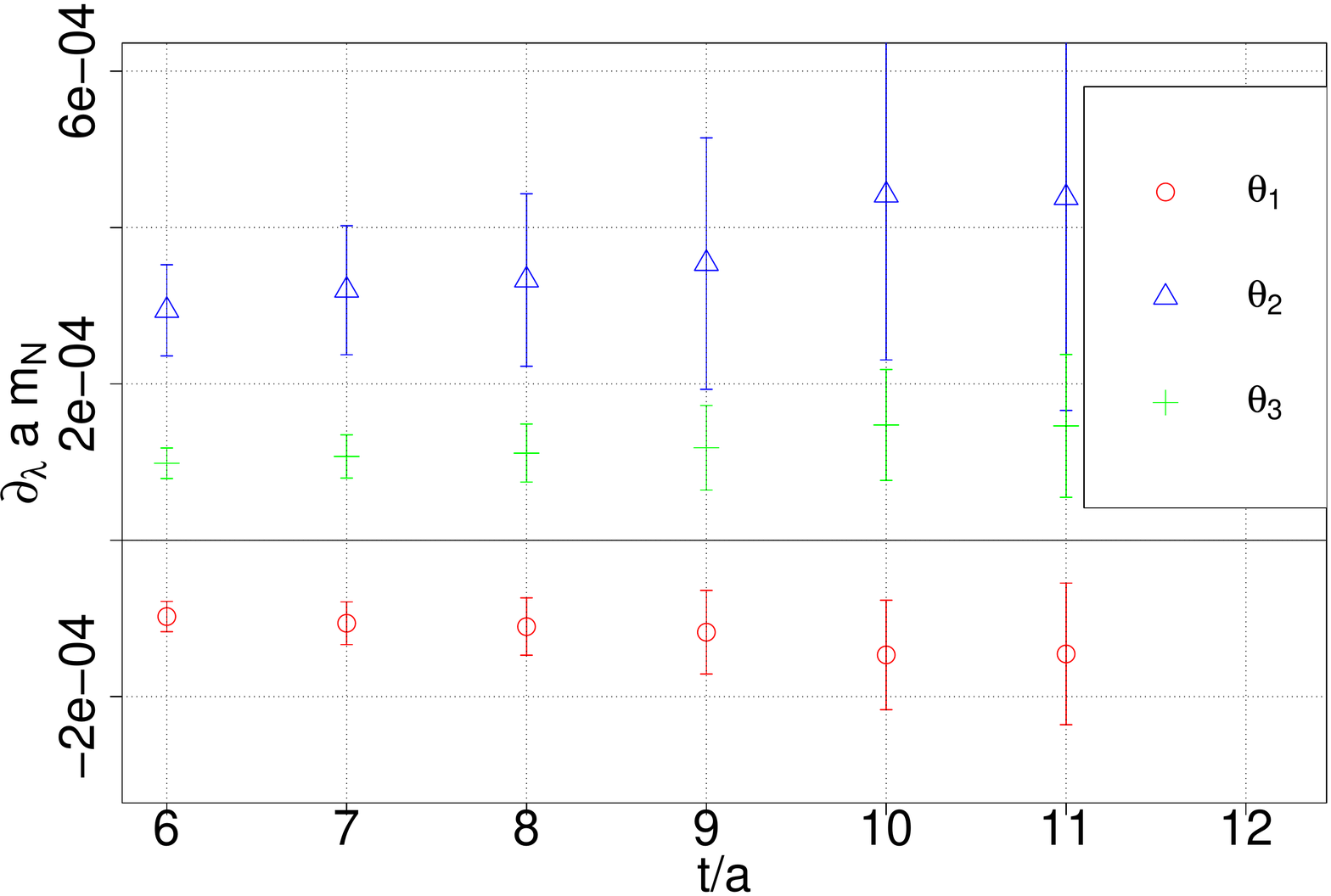}
  \caption{Mass shifts fitted over a range of $\tau/a \in [3,7]$ for each timeslice}
  \label{fig:dm_t}
\end{figure}

The mass shift is calculated according to FHT. Figure~\ref{fig:theta1_t_tau} shows the $t$ and $\tau$ dependence of the shift for $\theta_1$. The noise becomes prohibitively large for $t/a \geq 12$. Figure~\ref{fig:dm_t} shows the mass shift fitted in range $\tau / a \in [3,7]$ for the timeslices $t/a \in [6,11]$.

Given the exploratory nature of this study, we do not intend to
determine a value for the coupling $h_\pi^1$ here yet. However, the
discrepancy between the value obtained by naively adding these three
plateau values with corresponding Wilson coefficients and the
experimental value most likely stem from the missing renormalization
process and the missing strange contributions. We are currently
working on both.

\section{Conclusion}

This work represents an exploratory benchmark for a newly proposed
technique to measure the coupling $h_{\pi}^1$. So far the results 
are promising. Non-zero signal has been observed for all
diagrams with all Dirac structures. The mass shifts for individual
operators have also turned out to be non-zero. Thus the practical simplification of signal extraction envisioned by FGS is reachable.

Work on the strange operators and renormalization of the 4-quark lattice operators is currently ongoing. A parallel computation with traditional 3-point technique is also underway in order to compare cost and signal quality with the current work. 
Once our setup is optimized, we aim at a calculation directly at physical pion mass to obtain a theoretical estimate for $h_\pi^1$, which is
rigorously comparable to the experimental determination.

\section*{Acknowledgments}

This work is supported by the Deutsche
Forschungsgemeinschaft (DFG, German Research Foundation) and the  
NSFC through the funds provided to the Sino-German
Collaborative Research Center CRC 110 “Symmetries
and the Emergence of Structure in QCD” (DFG Project-ID 196253076 -
TRR 110, NSFC Grant No.~12070131001).

\end{document}